\documentclass[letterpaper, 10 pt, conference]{ieeeconf}  
\IEEEoverridecommandlockouts                              

\usepackage{url}
\usepackage{amsmath}
\usepackage{amssymb}
\usepackage[capitalize,noabbrev]{cleveref}
\usepackage{subcaption}

\usepackage{tikz-cd}
\usepackage{pgffor}

\usepackage{enumerate}

\usepackage[textsize=tiny]{todonotes}
\usepackage[dvipsnames]{xcolor}
\usepackage{ULiege-colors}

\usepackage[acronym, toc]{glossaries}
\newacronym{hpm}{HPM}{High-Pass Model}
\newacronym{cbm}{CBM}{Conductance-Based Model}
\newacronym{lif}{LIF}{Leaky Integrate-and-Fire}
\newacronym{hh}{HH}{Hodgkin-Huxley}
\newglossaryentry{sig}{
    name=$\sigma$,
    description={Sigmoid Function}
}
\makeglossaries
\newcommand{\rheoUp}{\ensuremath{\overline{I_{\mathrm{rheo}}}}}
\newcommand{\rheoLow}{\ensuremath{\underline{I_{\mathrm{rheo}}}}}

\usepackage[load-configurations=version-1]{siunitx}
\usepackage{tikz, pgfplots}
\pgfplotsset{
    compat=1.17,
    every axis/.append style={
      axis lines=left,
      width=8cm, height=5.5cm,
      grid=none, grid style={gray!20},
      legend cell align=left,
      legend style={font=\scriptsize, inner sep=1pt},
    },
}

\usetikzlibrary{arrows.meta, arrows, positioning, svg.path, shapes.geometric, calc, fadings, angles, quotes, backgrounds}
\usepgfplotslibrary{groupplots, statistics, patchplots}
\tikzstyle{block} = [draw, fill=white, rectangle, inner sep=0.5em]
\tikzstyle{blocknonlin} = [draw, fill=gray!20, rectangle, inner sep=0.5em]
\tikzstyle{sum} = [draw, circle]
\tikzstyle{connector} = [->, >=stealth', very thick]

\title{\vspace{2.0pt}\LARGE \bf
    Excitable Behaviours of a First-Order High-Pass Neuron Model
}

\renewcommand{\authorrefmark}[1]{\raisebox{0pt}[0pt][0pt]{\textsuperscript{\footnotesize #1}}}

\author{Jérôme Adriaens\authorrefmark{1}, Guillaume Drion\authorrefmark{1}, Pierre Sacré\authorrefmark{1}, Alessio Franci\authorrefmark{1,2}, Gustave Bainier\authorrefmark{1}%
\thanks{\authorrefmark{1}J.\ Adriaens, G.\ Drion, P.\ Sacré, A.\ Franci, and G.\ Bainier are with the Neuroengineering Lab, Department of Electrical Engineering and Computer Science, University of Liège, Belgium.
\authorrefmark{2}A.\ Franci is with the WEL-T Department, WEL Research Institute, 1300 Wavre, Belgium. 
 \{jadriaens, gdrion, p.sacre, afranci, gustave.bainier\}@uliege.be.}%
}

\colorlet{k_col}{ulgRed}
\colorlet{p_col}{ulgBlue}
\colorlet{alpha_col}{ulgGreen}

\usepackage{stfloats}

\let\labelindent\relax
\usepackage{enumitem}
\begin{document}

\maketitle
\thispagestyle{empty}
\pagestyle{empty}
\begin{abstract}
With the rise of neuromorphic technologies, simple and tunable neuron models are needed to analyse and design spiking systems. However, the simplest neuron models, e.g., Leaky Integrate-and-Fire (LIF) models, usually fail to capture the excitable behaviours of biological neurons. This could be a significant limitation, as excitability enables reliable spike timing under parametric uncertainties. Recently, a minimal excitable model, the so-called High-Pass Model (HPM), was introduced in neuroengineering. It simply consists of a static sigmoid with a high-pass filter in positive feedback. Given its novelty, it is still unclear how easily the HPM can be tuned, and how much of the excitable behaviours of a biophysical neuron it can reproduce. To address these questions, we investigate the tunability of a first-order HPM sub- and supra-rheobase, i.e., below and above the onset of the tonic spiking regime (where the neuron spikes periodically), and compare its behaviours with those of a simple Hodgkin–Huxley-type Conductance-Based Model (CBM). Below rheobase, both models fire spikes in response to transient inputs, including post-inhibitory rebound spikes. Above rheobase, an affine reparametrization of the input of the HPM suffices to fit its frequency-current curve to that of a CBM. The two models mostly differ near their respective rheobases, where the first-order HPM lacks bistability and spike latency. We suggest that modifying the frequency-selective element of the HPM may be sufficient to overcome this limitation. These results highlight the potential of the HPM as a fundamental computational primitive in neuromorphic, spike-based technologies.%
\end{abstract}

\section{Introduction}

Neurons are excitable cells: they emit spikes with reproducible timing in response to transient inputs~\cite{MainenSejnowski1995,sepulchreExcitableBehaviors2017}. Since Hodgkin and Huxley~\cite{HodgkinHuxley1952}, \glspl{cbm} have been the reference description of the ionic mechanisms behind excitability. However, dimensionality and nonlinearities of \glspl{cbm} make their dynamics, and in particular their spiking threshold, challenging to analyse and exploit in the design of neuromorphic, spike-based technologies.

While this complexity has motivated reduced models of \glspl{cbm} that retain the essence of their biophysics~\cite{DrionSepulchre2021}, the models mostly used in neuromorphic designs are from the integrate-and-fire family~\cite{Burkitt2006}, since they are very easy to analyse. In particular, a common choice is the linear \gls{lif} neuron. However, a \gls{lif} neuron acts as a low-pass system, and thus is non-excitable: it spikes exactly when its input crosses a static threshold, so its response depends only on the input amplitude, not on how rapidly the input changes. Excitable neurons, in contrast, have a dynamic spiking threshold that depends on both the amplitude and the rate of change of the input~\cite{AzouzGray2000,PlatkiewiczBrette2010}. Although extensions such as the adaptive exponential model recover some of these excitable behaviours~\cite{BretteGerstner2005,Izhikevich2000}, this is at the price of additional states, parameters and nonlinearities, reintroducing some of the complexity that \gls{lif} models seek to avoid.

\begin{figure*}[t]
    \centering
    \vspace{4pt}
%
\providecommand{\rasterxmin}{350}
\providecommand{\rasterxmax}{550}

\begin{tikzpicture}
\begin{groupplot}[
  group style={group name=rg, group size=2 by 2, horizontal sep=0.5cm, vertical sep=0.8cm, x descriptions at=edge bottom, y descriptions at=edge left},
  scale only axis, width=0.5\textwidth-1.0cm, height=1.cm,
  xmin=\rasterxmin, xmax=\rasterxmax, ymin=-1.5, ymax=50.5,
  ytick=\empty, tick align=outside, major tick length=2pt,
  /pgf/number format/1000 sep={},
  tick label style={font=\scriptsize}, label style={font=\footnotesize},
  xlabel style={yshift=2pt}, ylabel style={yshift=-2pt},
  title style={font=\footnotesize\bfseries, yshift=-4pt},
  xlabel={Time (ms)}, ylabel={},
]
\pgfplotsset{raster/.style={only marks, mark=|, mark size=0.3pt, line width=0.5pt, black}}

\nextgroupplot[title={\gls{cbm}}]
\addplot[raster] table[x=t_ms, y=trial] {_datas/reliability/raster_GNEU_II.dat};
\nextgroupplot[title={LIF}]
\addplot[raster] table[x=t_ms, y=trial] {_datas/reliability/raster_LIF.dat};
\nextgroupplot[title={\gls{hpm}}]
\addplot[raster] table[x=t_ms, y=trial] {_datas/reliability/raster_HPM.dat};
\nextgroupplot[title={Non-excitable \gls{hpm}}]
\addplot[raster] table[x=t_ms, y=trial] {_datas/reliability/raster_HPM_satk.dat};
\end{groupplot}

\node[anchor=south, align=center, fill=none, rotate=90, font=\footnotesize] at ($(rg c1r1.west)!0.5!(rg c1r2.west)$) {Neuron index};
\path let \p1=(rg c1r1.north west), \p2=(rg c1r1.north east) in
  \pgfextra{\pgfmathsetlengthmacro{\rasterw}{\x2-\x1}\global\let\rasterw\rasterw};
\begin{axis}[
  at={(rg c1r1.north west)}, anchor=south west, yshift=0.8cm,
  scale only axis, width=\rasterw, height=2.2cm,
  xmin=\rasterxmin, xmax=\rasterxmax, scaled x ticks=false,
  tick align=outside, major tick length=2pt, tick style={line width=0.3pt},
  xticklabels={}, ymin=-50, ymax=60, ytick=\empty,
  tick label style={font=\scriptsize}, label style={font=\footnotesize},
  ylabel={$I_\text{ext}$}, ylabel style={yshift=-2pt},
]
\addplot[black!70, line width=0.25pt] table[x=t_ms, y=I_total] {_datas/reliability/stimulus.dat};
\end{axis}

\begin{axis}[
  at={(rg c2r1.north west)}, anchor=south west, yshift=0.8cm,
  scale only axis, width=\rasterw, height=2.2cm,
  xmin=10, xmax=90, ymin=40, ymax=100, ytick={0,50,75},
  tick align=inside, major tick length=2pt, tick style={line width=0.3pt},
  tick label style={font=\scriptsize}, label style={font=\footnotesize},
  xlabel={$I_\text{ext}$}, ylabel={$f$ (Hz)},
  xlabel style={yshift=2pt}, ylabel style={yshift=-2pt},
  legend style={font=\tiny, draw=none, fill=none, at={(0.6,0.05)}, anchor=south east,
    inner sep=1pt, row sep=-2pt},
  legend columns = 2,
  legend cell align=left,
  xlabel style={at={(axis description cs:1,0)}, anchor=south east},
  ylabel style={at={(axis description cs:0,1.0)}, rotate=-90, anchor=north west},
]
\pgfplotsset{fi/.style={line width=2.5pt, restrict x to domain=12.15:100}}
\addplot[fi, gray]     table[x=I, y=f] {_datas/reliability/fi_GNEU_II.dat};
\addlegendentry{\gls{cbm}}
\addplot[fi, ulgOrange]   table[x=I, y=f] {_datas/reliability/fi_HPM_satk.dat};
\addlegendentry{Non-excitable \gls{hpm}}
\addplot[fi, ulgBlue, densely dashed] table[x=I, y=f] {_datas/reliability/fi_HPM.dat};
\addlegendentry{\gls{hpm}}
\addplot[fi, ulgRed, loosely dashed]    table[x=I, y=f] {_datas/reliability/fi_LIF.dat};
\addlegendentry{LIF}
\end{axis}
\end{tikzpicture}
    \caption{\small \textbf{Parametric robustness in spike timing of excitable models.} Raster of the spike times of 50 copies of each model (one row per copy), whose parameters are perturbed by a 5\% relative standard deviation noise, driven by the same fluctuating current (top left). The initial models are chosen to share a similar tonic spiking regime (top right). The \gls{cbm} and the \gls{hpm} fire at nearly the same times across copies (left). The non-excitable \gls{hpm} and the \gls{lif} do not (right). All code and data are available on GitHub.}
    \label{fig:reliability-exp}
\end{figure*}
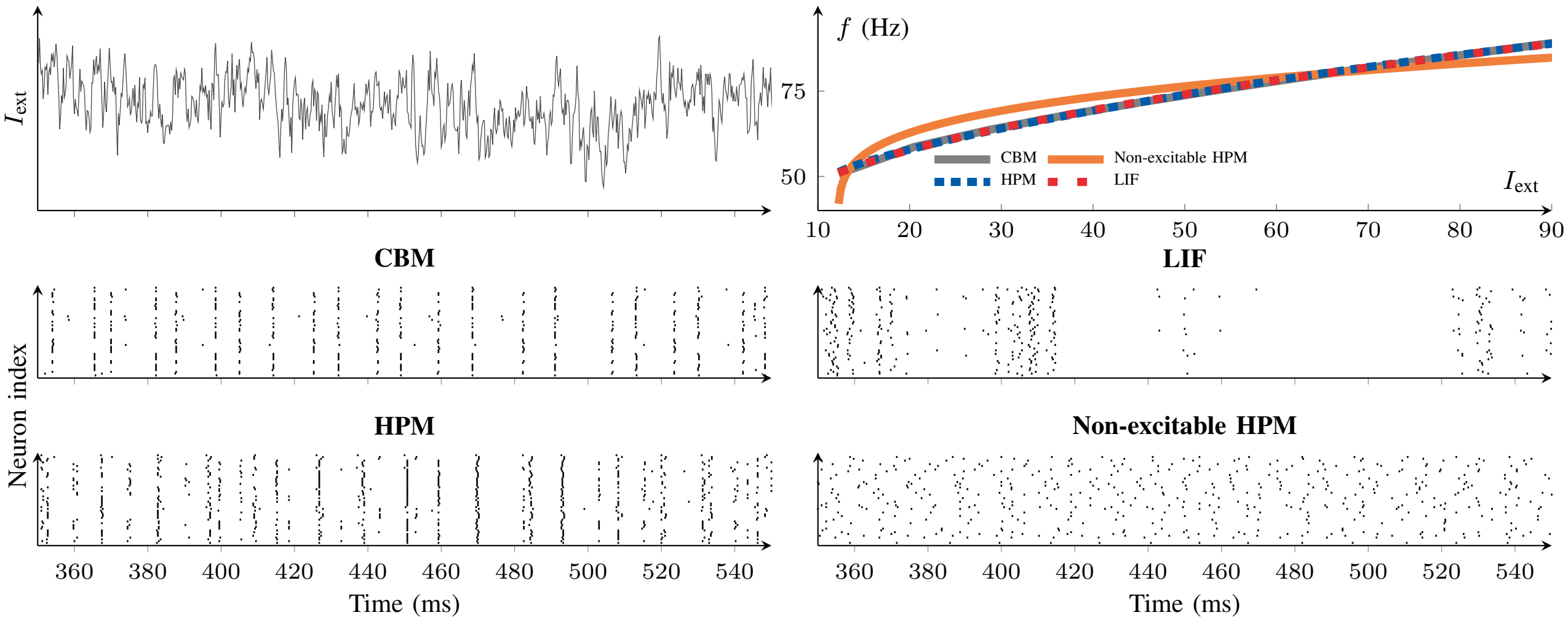

Excitability is, however, at the heart of the spike timing robustness of neurons. Under the same fluctuating input, perturbed copies of excitable models such as \glspl{cbm} spike more reliably than non-excitable models (\cref{fig:reliability-exp}). While the \gls{lif} also shows some amount of reliability, its spike timings are also less consistent, and, crucially, its response is qualitatively different from that of a \gls{cbm}. This suggests that neuromorphic technologies could benefit from easily tunable and theoretically tractable spiking primitives that retain neuronal excitability, particularly for bio-inspired spike-based coding in the presence of hardware mismatch. 

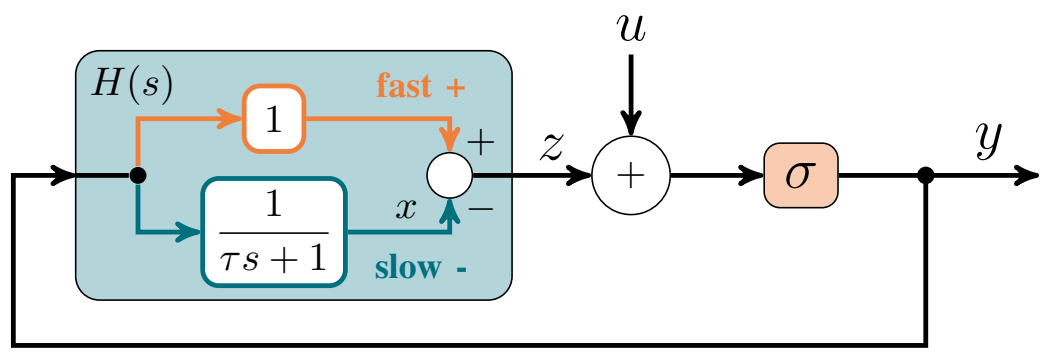
\begin{figure}[t]
    \centering
    \resizebox{0.8\columnwidth}{!}{
\begin{tikzpicture}[auto, transform shape]
    \draw [rounded corners=6pt, fill=ulgBlueGreen!30] (-0.3,-1.1) rectangle (3.55,1.1);
    \node [align=center, font=\bfseries, anchor=north west] at (-0.3,1.1) {$H(s)$};
    \coordinate (hin) at (-0.3,0);
    \node [circle, fill=black, inner sep=1.5pt] (split) at (0.25,0) {};
    \node [block, draw=ulgOrange, very thick, rounded corners=4pt] (one) at (1.45,0.5) {$1$};
    \node [block, draw=ulgBlueGreen, very thick, inner sep=0.3em, rounded corners=4pt] (lp) at (1.45,-0.5) {$\dfrac{1}{\tau s+1}$};
    \node [sum, fill=white, minimum size=0.4cm, inner sep=0pt] (diff) at (3.0,0) {};

    \draw [very thick] (hin) -- (split);
    \draw [connector, ulgOrange] (split.north) |- (one.west);
    \draw [connector, ulgOrange] (one.east) -| node [pos=0.85, right, black] {$+$} (diff.north);
    \draw [connector, ulgBlueGreen] (split.south) |- (lp.west);
    \draw [connector, ulgBlueGreen] (lp.east) -| node [pos=0.85, right, black] {$-$} (diff.south);
    \node [ulgOrange, font=\small\bfseries] at (2.75,0.8) {fast +};
    \node [ulgBlueGreen, font=\small\bfseries] at (2.75,-0.8) {slow -};
    \node at (2.62,-0.3) {$x$};

    \node [sum] (sum2) at (4.6,0) {$+$};
    \node (map) at (4.6,1.3) {\Large $u$};
    \node [blocknonlin, fill=ulgOrange!40, rounded corners=4pt] (nonlin) at (6.1,0) {\Large $\sigma$};
    \coordinate (branch) at (7.2,0);
    \coordinate (end) at (8.2,0);

    \draw [very thick] (diff.east) -- (3.55,0);
    \draw [connector] (3.55,0) -- node [midway, anchor=south] {\Large $z$} (sum2.west);
    \draw [connector] (map.south) -- (sum2.north);
    \draw [connector] (sum2.east) -- (nonlin.west);
    \draw [connector] (nonlin.east) -- node [above, near end] {\Large $y$} (end);
    \fill (branch) circle (2.2pt);
    \draw [connector] (branch) -- (7.2,-1.5) -- (-0.85,-1.5) -- (-0.85,0) -- (hin);
\end{tikzpicture}}
    \caption{\small \textbf{The first-order high-pass neuron model.}} 
    \label{fig:HPM}
\end{figure}

Recently, a minimal excitable model, the so-called \gls{hpm}, was introduced in neuroengineering~\cite{Bainier2026_poster}. It consists of a static high-gain sigmoid with a high-pass filter in positive feedback. According to the mixed-feedback principle, excitability in biophysical neurons results from the interplay of a fast positive and a slow negative feedback~\cite{DrionFranciDethierSepulchre2015,SepulchreDrionFranci2019}. The first-order \gls{hpm} might be the simplest realization of this principle available: the feedthrough branch of the high-pass acts as a fast positive feedback, while the low-pass branch acts as a slow negative feedback~\cite{Bainier2026_poster} (\cref{fig:HPM}). Like other excitable models, the \gls{hpm} can generate spikes in response to sub-rheobase transient inputs and produce post-inhibitory rebound spikes (where a spike is produced at the release of an inhibitory input). It also exhibits robust spike timing (\cref{fig:reliability-exp}). In short, the \gls{hpm} is a neuron model that appears to reproduce the excitable behaviours of a \gls{cbm} while retaining minimal model complexity.

This raises two practical questions: how expressive and easily tunable is the \gls{hpm}? In particular, can it be tuned to reproduce the behaviours of a simple \gls{cbm}? The answers will clarify which neuronal behaviours the \gls{hpm} can easily support for neuromorphic applications, and which properties of a \gls{cbm} can be captured and analysed with a \gls{hpm}.
We address these questions using a simple Hodgkin-Huxley-type \gls{cbm} as a reference, and we deliberately keep the \gls{hpm} as simple as possible: the high-pass filter is fixed to first-order.
Our contributions are the following.
\begin{enumerate}
    \item Below the rheobase, we show that the \gls{hpm} reproduces the main excitable behaviours of the \gls{cbm}, and that its parameters have a clear and intuitive influence on its input sensitivity.
    \item Above the rheobase, we show that the frequency-to-input-current ($f$-$I$) curve of the \gls{cbm} can be matched by the \gls{hpm} using an analytical least-squares fit and an affine parametrization of its input.
    \item We show that the remaining differences between the \gls{hpm} and the simple \gls{cbm} concern specific mechanisms of the \gls{cbm} near its rheobase: the \gls{cbm} possesses a noticeable bistability region~\cite{Hill1936,AzouzGray2000}, and it can significantly delay its spike initiation after a slow excitation. In contrast, the first-order \gls{hpm} is monostable and generally fires immediately after excitation.\label{item:third-point}
\end{enumerate}

This work comes with three accompanying papers. In \cite{BAINIER-ECC}, the excitability mechanism of the \gls{hpm} is introduced, together with an explicit dynamical threshold separating the resting and excited states. In \cite{CHRI-JE-ECC}, an analytical expression for the tonic spiking frequency of the \gls{hpm} is obtained. Finally, in \cite{ARTHUR-ECC}, both parabolic and elliptic bursting are reproduced with minimal modifications to the \gls{hpm}. Collectively, these results establish a set of design principles for the use of the \gls{hpm} as a neuromorphic computational primitive.

The remainder of the paper is organized as follows. \Cref{sec:models} introduces the \gls{hpm} and \gls{cbm} models. \cref{sec:excitability} compares their responses to transient inputs and how the \gls{hpm} can be tuned. \cref{sec:tonic_regime} compares their tonic spiking regimes under constant inputs and fits the \gls{hpm} $f$-$I$ curve. Finally, \cref{sec:conclusion} discusses the limitations of first-order \glspl{hpm} and future directions.
All code and data are available on GitHub: \url{https://github.com/JeromeAdriaens/EB-FO-HPM-ECC27}.

\section{Hodgkin-Huxley and high-pass models}\label{sec:models}

This section introduces the two models compared in this paper: a Hodgkin-Huxley \gls{cbm}, used as a simple but representative biological neuron model, and a first-order \gls{hpm}. While the \gls{cbm} is rooted in the biophysical, circuit-level implementation of excitability, the \gls{hpm} is derived from teleological, i.e., function-driven, principles, retaining only the minimal input-output elements underlying excitability. In both models, tonic spiking emerges when their input crosses the rheobase, a point at which their resting equilibrium bifurcates into a limit cycle. By contrast, a neuron is excitable if a transient input, possibly below the rheobase, can still trigger a spike. Both models follow a mixed-feedback structure and are excitable.

\subsection{A Hodgkin-Huxley conductance-based model}

We consider a Hodgkin-Huxley \gls{cbm} with a usual minimal set of ionic currents for action potential generation~\cite{HodgkinHuxley1952}: a leak current $I_\mathrm{l}$, a transient sodium current $I_\mathrm{Na}$, and a delayed-rectifier potassium current $I_\mathrm{K,d}$. With an applied current $I_{\mathrm{ext}}$, the membrane potential obeys Kirchhoff's current balance: $C\dot{V}=-\sum I_{\text{ion}}+I_{\mathrm{ext}}$, i.e.
\begin{equation}
    \begin{split}
    C \dot V = &- \bar{G}_{\mathrm{l}}(V-E_{\mathrm{l}}) - G_{\mathrm{Na}}(V,t)(V-E_{\mathrm{Na}})\\
    &- G_{\mathrm{K,d}}(V,t)(V-E_{\mathrm{K}}) + I_{\mathrm{ext}}(t),
    \end{split}
    \label{eq:hh_model}
\end{equation}
with $G_{\mathrm{Na}}=\bar{G}_{\mathrm{Na}}\,m_{\mathrm{Na}}h_{\mathrm{Na}}$ and $G_{\mathrm{K,d}}=\bar{G}_{\mathrm{K,d}}\,m_{\mathrm{K,d}}$ the dynamic conductances. Each gating variable $X\in\{m_{\mathrm{Na}},h_{\mathrm{Na}},m_{\mathrm{K,d}}\}$ obeys
\begin{equation}
    \tau_{X}(V)\dot X = X_{\infty}(V)-X,
    \label{eq:gating}
\end{equation}
where $X_{\infty}(V;a,b)=\big(1+\exp\frac{V+a}{b}\big)^{-1}$ is a Boltzmann function and $\tau_{X}(V;a,b,d,e)=a-b\cdot(1+\exp\frac{V+d}{e})^{-1}$ a shifted Boltzmann function. Contrary to the original Hodgkin-Huxley model~\cite{HodgkinHuxley1952}, where the gating variables are raised to higher powers to fit the kinetics of the ion channels precisely, here they enter with exponent one, since we only care about the qualitative structure of the model.
In this model, sodium activation $m_{\mathrm{Na}}$ provides a fast positive feedback on $V$, whereas potassium activation $m_{\mathrm{K,d}}$ and sodium inactivation $h_{\mathrm{Na}}$ provide slow negative feedbacks, so the system is excitable and fits the mixed-feedback principle~\cite{DrionFranciDethierSepulchre2015,SepulchreDrionFranci2019}. 

\subsection{The first-order high-pass neuron model}
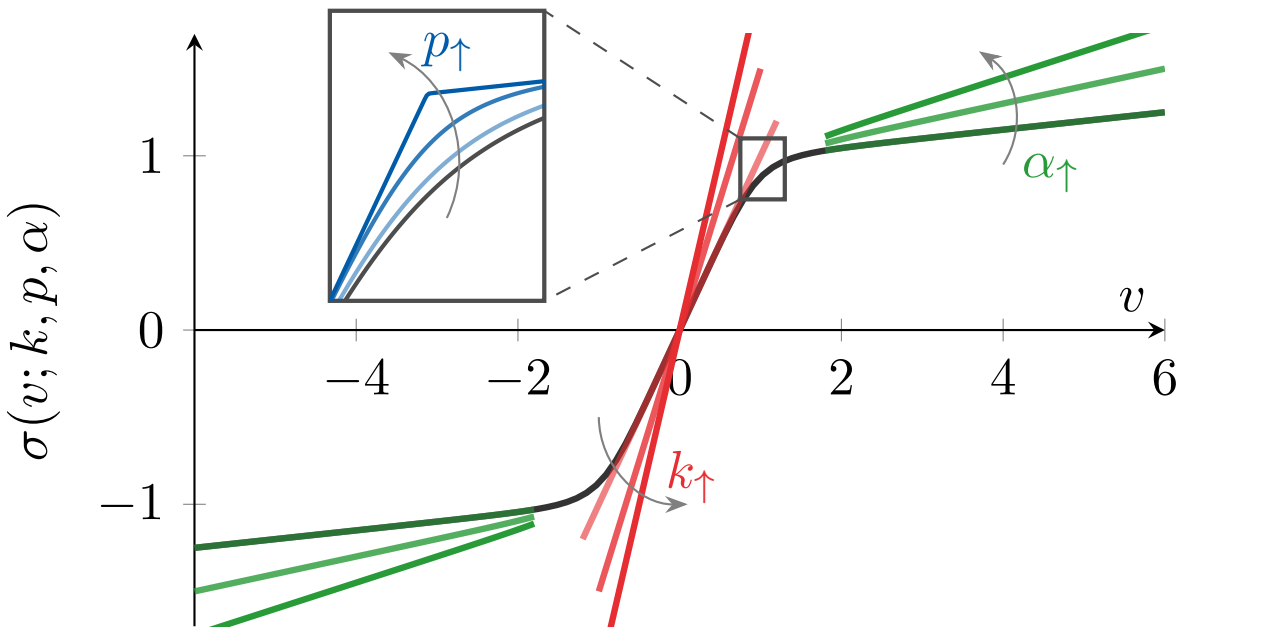
\begin{figure}[t]
    \centering
    \vspace{5pt}
    \begin{tikzpicture}
\tikzset{
    declare function={
        sigmakp(\v,\k,\p,\a) = \a*\v+(1-(\a/\k))*((\k*\v)/(1+abs(\k*\v)^\p)^(1/\p));
    }
}
\pgfplotsset{
    every axis plot/.append style={very thick},
    }
\begin{axis}[
    axis lines = left,
    xlabel = {$v$},
    ylabel = {$\sigma(v;k,p,\alpha)$},
    ymax = 1.7,
    ymin=-1.7,
    domain = -6:6,
    samples = 100,
    no markers,
    axis x line=center,
    legend style={at={(0.5,1.2)}, draw=none, fill=none, anchor=north},
    legend columns=3,
]
    \addplot[black!80] {sigmakp(x, 1, 5, 0.05)};
    
    \addplot[k_col, domain=-.5:.5, opacity=0.6] coordinates {(-1.2,-1.2) (0,0) (1.2,1.2)};
    \addplot[k_col, domain=-.5:.5, opacity=0.8] coordinates {(-1,-1.5) (0,0) (1,1.5)};
    \addplot[k_col, domain=-.5:.5, ] coordinates {(-1,-2.0) (0,0) (1,2.0)};
    \draw[-{Stealth[length=1.5mm]}, gray, bend left=-45, thin]
        (axis cs:-1.,-0.5) to (axis cs:0.1,-1.0);
    \node[font=\normalsize, anchor=south, k_col] at (axis cs:0.15,-1.1) {$k_\uparrow$};
    
    \foreach \xa/\xb in {1.8/6, -6/-1.8}{
        \foreach \opc/\aval in {0.6/0.05, 0.8/0.1, 1.0/0.15}{
            \expanded{\noexpand\addplot[alpha_col, domain=\xa:\xb, opacity=\opc] {sigmakp(x, 1, 5, \aval)};}
        }
    }
    \draw[-{Stealth[length=1.5mm]}, gray, bend left=-45, thin]
        (axis cs:4.,0.95) to (axis cs:3.7,1.6);
    \node[font=\normalsize, anchor=west, alpha_col] at (axis cs:4.0,0.9) {$\alpha_\uparrow$};
    
    \draw[black!70, thick] (axis cs:0.75,0.75) rectangle (axis cs:1.3,1.1);
    \coordinate (zoomBL) at (axis cs:0.75,0.75);
    \coordinate (zoomTL) at (axis cs:0.75,1.1);
    \coordinate (insetpos) at (axis cs:-3,1.);   
\end{axis}

\begin{axis}[
    name=zoom,
    at={(insetpos)}, anchor=center,
    width=3cm, height=3.5cm,
    xmin=0.75, xmax=1.3, ymin=0.75, ymax=1.1,
    domain=0.7:1.8, samples=200, no markers,
    axis lines=box,
    xtick={0}, ytick={0},
    tick label style={font=\tiny},
    every axis plot/.append style={thick},
    axis background/.style={fill=white},
    axis line style={black!70, thick},
]
    \addplot[black!70]   {sigmakp(x, 1, 5, 0.05)};
    \addplot[p_col, opacity=0.5]        {sigmakp(x, 1, 6, 0.05)};
    \addplot[p_col, opacity=0.8]        {sigmakp(x, 1, 10, 0.05)};
    \addplot[p_col]        {sigmakp(x, 1, 500, 0.05)};
    
    \draw[-{Stealth[length=1.5mm]}, gray, bend left=-45, thin]
        (axis cs:1.05,0.85) to (axis cs:0.9,1.05);
    \node[font=\normalsize, anchor=west, p_col] at (axis cs:0.94,1.05) {$p_\uparrow$};
    
\end{axis}

\draw[black!70, dashed] (zoomBL) -- (zoom.south east);
\draw[black!70, dashed] (zoomTL) -- (zoom.north east);
\end{tikzpicture}
    \caption{\small \textbf{Parametrized sigmoid function.} Arrows show the effect of increasing $k$ (\textcolor{ulgRed}{red}), $p$ (\textcolor{ulgBlue}{blue}), and $\alpha$ (\textcolor{ulgGreen}{green}).}
    \label{fig:sigmoid-theory}
\end{figure}

The \gls{hpm} was introduced in~\cite{Bainier2026_poster}. It consists of a static high-gain sigmoid $\sigma$ with a first-order high-pass filter $H(s)$ in positive feedback. At first-order, $H(s):=\tau s/(\tau s+1)$ (\cref{fig:HPM}). In state-space form, the model reads
\begin{subequations}  \label{eq:hpm}
\begin{align}
    y&=\sigma(u+y-x),\\
     \tau\dot x&=-x+y,
\end{align}
\end{subequations}
with $u$ the input signal, $y$ the output signal, $x$ the internal state of high-pass. We moreover define $v:=u+y-x$ as the sigmoid input. Since $y$ appears on both sides of $y=\sigma(u+y-x)$, the motif is acausal. To simulate it, we use the causal regularization $\varepsilon \dot y = -y+\sigma(u+y-x)$. The state-space equations \eqref{eq:hpm} are recovered at the singular limit $\varepsilon \to 0^+$, corresponding to the classical singularly-perturbed limit of neurodynamics.

The first-order high-pass combines an instantaneous positive feedback with a slow negative feedback, which fits the mixed-feedback principle. The model can also be understood as a lead-excitation motif: the high-pass filter computes the trend of the output (lead), which is then amplified by positive feedback (excitation). Collapsing the feedthrough branch of the high-pass filter with the sigmoid yields a bistable block, so the system can also be read as a hysteresis-and-adaptation motif, much like the FitzHugh-Nagumo neuron model~\cite{BAINIER-ECC}.

The tunability of the first-order \gls{hpm}, beyond the high-pass timescale $\tau$, lies entirely in the choice of the high-gain sigmoid $\sigma$, which we parametrize as follows
\begin{equation}
    \sigma(v;{\color{k_col}k},{\color{p_col}p},{\color{alpha_col}\alpha}) = {\color{alpha_col}\alpha}\cdot v + ({\color{k_col}k}-{\color{alpha_col}\alpha})\cdot\cfrac{v}{\left(1+|{\color{k_col}k}\cdot v|^{\color{p_col}p}\right)^{{\color{p_col}p}^{-1}}}.
\end{equation}
The effects of the three parameters ($k>1$, $p>0$ and $\alpha \geq 0$) on the sigmoid are illustrated in~\cref{fig:sigmoid-theory}. The slope at the origin is $\sigma'(0)=k$, so the corners of $\sigma$ lie near $(\pm\frac{1}{k},\pm 1)$. The exponent $p$ sets the sharpness of these corners, so $\sigma'$ becomes discontinuous at $\pm\frac{1}{k}$ as $p\to\infty$. Finally, $\alpha$ is the slope of $\sigma$ as $v\to\pm\infty$. The rheobase of the model lies at the point where $\sigma'=1$~\cite{CHRI-JE-ECC,BAINIER-ECC,ARTHUR-ECC}. Consequently, $\alpha$, $p$ and $k$ set the input sensitivity at rest, near the rheobase, and during tonic spiking, respectively. Setting $(p,\alpha)=(\infty,0)$ yields a non-excitable \gls{hpm} neuron, since only inputs above the rheobase affect the neuron.

The rest of the paper studies the input-output behaviours of the first-order \gls{hpm}~\eqref{eq:hpm}, compares them to those of the \gls{cbm}~\eqref{eq:hh_model}, and examines how the parameters $\tau$, $k$, $p$, and $\alpha$ can shape them.

\begin{figure}[t]
    \centering
    \vspace{3pt}

\colorlet{spk0}{white}   \colorlet{spk1}{black!14}

\pgfplotsset{
  spkmap/.style={
    colormap={spkcm}{
      color(0pt)=(spk0)   color(0.99pt)=(spk0)
      color(1pt)=(spk1)   color(2pt)=(spk1)},
    point meta min=0, point meta max=2,
  },
  rheoStyle/.style={dashed, thick, rheoCol},
  contourLine/.style={black!70, thin},   
  wavemap/.style={width=6.4cm, height=5.6cm, enlargelimits=false, axis on top, axis background/.style={fill=spk0},
                  },
}

\tikzset{
  hlStyle/.style={very thick, hlCol},   
  hlLabel/.style={font=\scriptsize\bfseries, text=hlCol, fill=white, fill opacity=0.85,
                  text opacity=1, inner sep=1.5pt, anchor=north east},   
}

\providecommand{\hlregion}{}
\renewcommand{\hlregion}[6][hlCol]{%
  \draw[hlStyle, draw=#1] (axis cs:#2,#4) rectangle (axis cs:#3,#5);
  \node[hlLabel, text=#1] at (axis cs:#2,#5) {#6};}
\ifdefined\iffiHL\else\newif\iffiHL\fiHLtrue\fi

\def\fiPlotFills#1#2{\foreach \k in #2 {%
  \edef\fiTmp{\noexpand\addplot[fill=spk\k, draw=none, even odd rule]
    table[x=x, y=y] {_datas/alpha_sweep_theo_neg/#1_fill_\k.dat};}\fiTmp}}
\def\fiPlotLines#1#2{\foreach \k in #2 {%
  \edef\fiTmp{\noexpand\addplot[contourLine]
    table[x=x, y=y] {_datas/alpha_sweep_theo_neg/#1_line_\k.dat};}\fiTmp}}

\input{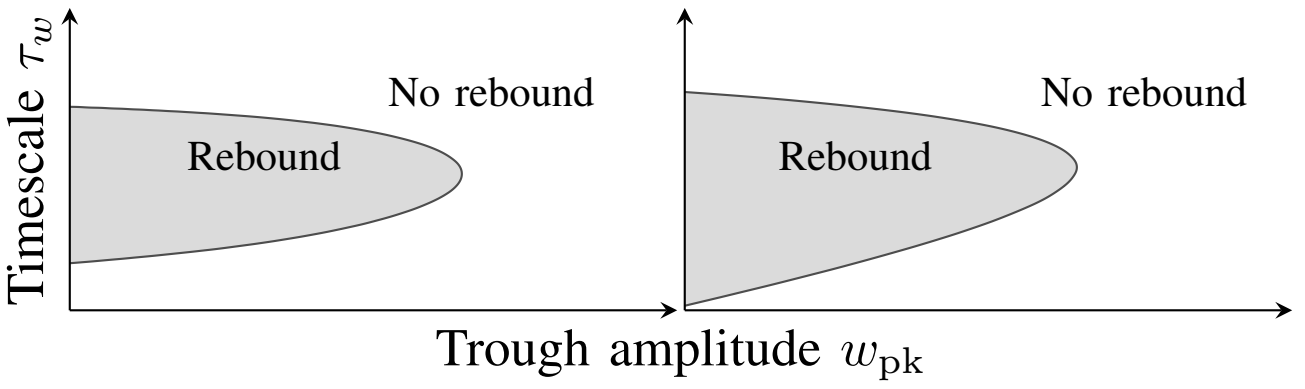}
\ifdefined\iffiRefLine\else\newif\iffiRefLine\fiRefLinetrue\fi

\begin{tikzpicture}[transform shape]
  \begin{groupplot}[
      group style={group size=2 by 1, horizontal sep=0.05cm},
      wavemap, spkmap, height=2cm,
      width=0.5\linewidth-0.3cm,
      x axis line style={stealth-},
      xmode=log, x dir=reverse, 
      scale only axis, ticks=none
    ]

    \nextgroupplot[
        xmin=\fiXmingneuii, xmax=\fiXmaxgneuii,
        ymode=log,
        axis y line=left,
        yticklabels={},
        ylabel={},
        ymin=\fiYmingneuii, 
        ymax=1.1e1, 
        xlabel={},
        xtick={}, 
        xticklabels={},
        ylabel={Timescale $\tau_w$}
      ]
      \fiPlotFills{gneu_ii}{\fiLevelsgneuii}
      \fiPlotLines{gneu_ii}{\fiLevelsgneuii}
      \node[text=black, rotate=0, font=\footnotesize] at (axis cs:30,3) {No rebound};
      \node[text=black, rotate=0, font=\footnotesize] at (axis cs:100,1.1) {Rebound};
    \nextgroupplot[ 
        xmin=125, 
        xmax=\fiXmaxals,
        ymode=log, ymin=1e-1, 
        ymax=1.1e1, 
        axis y line=left, 
        ylabel={},
        xlabel={},
        xtick={}, xticklabels={}, ytick={}
      ]
      \addplot[scatter, only marks, mark size=0pt, draw=none, point meta=explicit]
        coordinates {(\fiXminals,\fiYminals) [0]};
      \fiPlotFills{als}{\fiLevelsals}
      \fiPlotLines{als}{\fiLevelsals}
      \node[text=black, rotate=0, font=\footnotesize] at (axis cs:200,3.) {No rebound};
      \node[text=black, rotate=0, font=\footnotesize] at (axis cs:500,1.1) {Rebound};

  \end{groupplot}

  \node[anchor=north, align=center, fill=none]
    at ($(group c1r1.south)!0.5!(group c2r1.south)$) {Trough amplitude $w_{\mathrm{pk}}$};
  \node[anchor=north, text width=2cm, align=center, text height=0.1em] at ([yshift=-0.2cm]group c1r1.south) {\subcaption{\gls{cbm}.\label{fig:waveneg-gneu}}};
  \node[anchor=north, text width=2cm, align=center, text height=0.1em] at ([yshift=-0.2cm]group c2r1.south) {\subcaption{\gls{hpm}.\label{fig:waveneg-als}}};
\end{tikzpicture}
    \vspace{-15pt}
    \caption{\small \textbf{Rebound spikes} triggered by inhibitory waves~\eqref{eq:alpha_wave}, log-log scale. Non-excitable HPM and LIF do not exhibit rebound excitability.}
    \label{fig:alphawave_sweep_theory_rebound}
\end{figure}

\begin{figure*}[t]
    \raggedright
    \vspace{10pt}
%

\colorlet{spk0}{white}   \colorlet{spk1}{black!14}  \colorlet{spk2}{black!23}
\colorlet{spk3}{black!32}  \colorlet{spk4}{black!41}  \colorlet{spk5}{black!50}
\colorlet{spk6}{black!62}  \colorlet{spk7}{black!77}  \colorlet{spk8}{black!92}
\colorlet{rheoCol}{ulgBlueGreen}
\definecolor{hlCol}{RGB}{214,39,40}      
\colorlet{linkCol}{ulgBlueGreen}

\pgfplotsset{
  spkmap/.style={
    colormap={spkcm}{
      color(0pt)=(spk0)   color(0.99pt)=(spk0)
      color(1pt)=(spk1)   color(1.99pt)=(spk1)
      color(2pt)=(spk2)   color(2.99pt)=(spk2)
      color(3pt)=(spk3)   color(3.99pt)=(spk3)
      color(4pt)=(spk4)   color(4.99pt)=(spk4)
      color(5pt)=(spk5)   color(5.99pt)=(spk5)
      color(6pt)=(spk6)   color(6.99pt)=(spk6)
      color(7pt)=(spk7)   color(7.99pt)=(spk7)
      color(8pt)=(spk8)   color(9pt)=(spk8)},
    point meta min=0, point meta max=9,
  },
  rheoStyle/.style={very thick, rheoCol},
  contourLine/.style={black!70, thin},   
  wavemap/.style={width=6.4cm, height=5.6cm, enlargelimits=false, axis on top, axis background/.style={fill=spk0},
                  },
}

\tikzset{
  hlStyle/.style={very thick, hlCol},   
  hlLabel/.style={font=\scriptsize\bfseries, fill=none, fill opacity=0.85,
                  text opacity=1, inner sep=1.5pt, anchor=west, outer sep=0.0pt},
}

\providecommand{\hlregion}{}
\renewcommand{\hlregion}[6][hlCol]{%
  \draw[hlStyle, draw=#1, fill=#1, fill opacity=0.2] (axis cs:#2,#4) rectangle (axis cs:#3,#5);
  \node[hlLabel, text=#1, fill=none] at (axis cs:#2,#5) {#6};}
\ifdefined\iffiHL\else\newif\iffiHL\fiHLtrue\fi

\def\fiPlotFills#1#2{\foreach \k in #2 {%
  \edef\fiTmp{\noexpand\addplot[fill=spk\k, draw=none, even odd rule]
    table[x=x, y=y] {./_datas/alpha_sweep_theo/#1_fill_\k.dat};}\fiTmp}}
\def\fiPlotLines#1#2{\foreach \k in #2 {%
  \edef\fiTmp{\noexpand\addplot[contourLine]
    table[x=x, y=y] {./_datas/alpha_sweep_theo/#1_line_\k.dat};}\fiTmp}}

\input{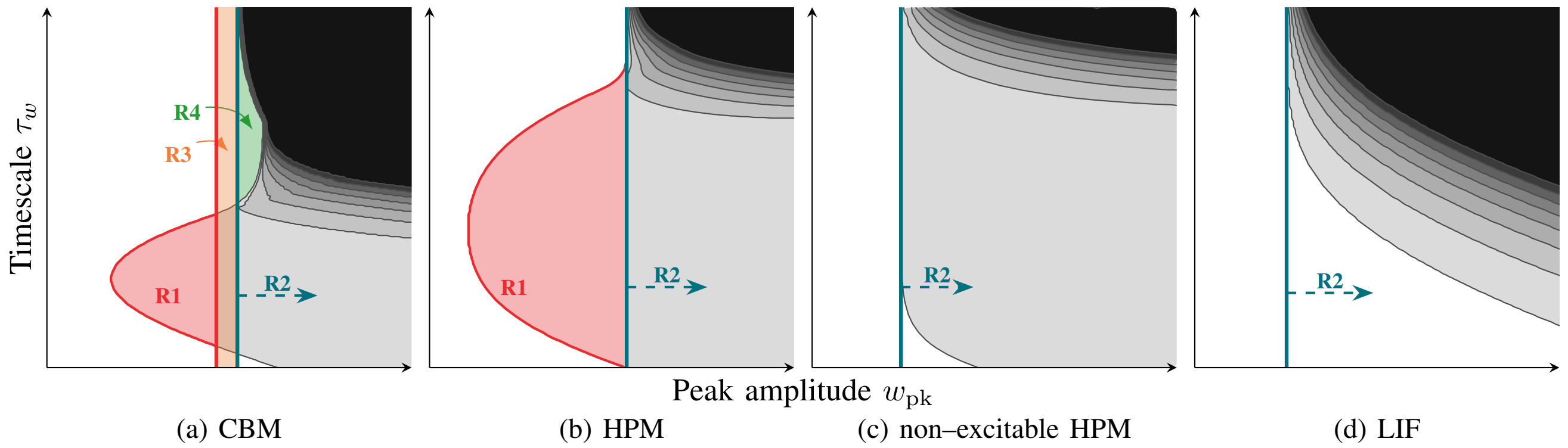}

\begin{tikzpicture}
  \begin{groupplot}[
      group style={group size=4 by 1, horizontal sep=0.2cm},
      wavemap, spkmap,
      height=4cm,
      ticks=none,
      y tick label style={font=\tiny},
      width=0.25\linewidth - 2*0.2cm,
      scale only axis
    ]

    \nextgroupplot[
        xmode=log, xmin=\fiXmingneuii, xmax=\fiXmaxgneuii,
        ymode=log, ymin=0.1, ymax=10000,, ylabel={Timescale $\tau_w$},
        xtick={5,10,20,40}, xticklabels={5,10,20,40},
        yticklabels={},
      ]
      \fill[ulgGreen, opacity=0.35] (axis cs:13.01, 10.0) rectangle (axis cs:17.0, 30000);
      \node[hlLabel, ulgGreen, anchor=west, name=HH_R4] at (axis cs: 8., 350.0) {R4};
      \draw[ulgGreen, -Latex] (HH_R4) to[bend left=30] (axis cs: 14.5, 200.);
      
      \fiPlotFills{gneu_ii_full}{\fiLevelsgneuii}
      \fiPlotLines{gneu_ii_full}{\fiLevelsgneuii}
      
      \fill[ulgOrange, opacity=0.35] (axis cs:11.207, 0.001) rectangle (axis cs:13.01, 30000);
      \node[ulgOrange, anchor=west, hlLabel, name=HH_R3] at (axis cs: 7.5, 90.0) {R3};
      \draw[ulgOrange, -Latex] (HH_R3) to[bend left=30] (axis cs: 12.2, 100.);
       
      \begin{scope}
          \clip (axis cs:4.0,0.1) rectangle (axis cs:11.207,50);
          \addplot[fill=ulgRed!35, draw=ulgRed, even odd rule, thick] table[x=x, y=y] {./_datas/alpha_sweep_theo/gneu_ii_full_fill_1.dat};
          \node[ulgRed, anchor=west, hlLabel] at (axis cs: 7, 1) {R1};
      \end{scope}
      \addplot[very thick, ulgRed] coordinates {(\fiRheogneuii,\fiYmingneuii) (\fiRheogneuii,10000)};
      \addplot[very thick, ulgBlueGreen] coordinates {(13.012,\fiYmingneuii) (13.012,10000)};
      \draw[thick, ulgBlueGreen, -Stealth, dashed] (axis cs: 13.01, 1) -- node[midway, hlLabel, ulgBlueGreen, anchor=south] {R2} (axis cs:23, 1);

    \nextgroupplot[
        xmode=log, xmin=0.19, xmax=\fiXmaxals,
        ymode=log, ymin=\fiYminals, ymax=\fiYmaxals, 
        xtick={0.2,0.4,0.8}, xticklabels={0.2,0.4,0.8},
        yticklabels={},
        ytick={},
      ]
      \addplot[scatter, only marks, mark size=0pt, draw=none, point meta=explicit]
        coordinates {(\fiXminals,\fiYminals) [0]};
      \fiPlotFills{als}{\fiLevelsals}
      \fiPlotLines{als}{\fiLevelsals}
      \begin{scope}
          \clip (axis cs:0.2,0.001) rectangle (axis cs:\fiRheoals,100);
          \addplot[fill=ulgRed!35, draw=ulgRed, even odd rule, thick] table[x=x, y=y] {./_datas/alpha_sweep_theo/als_fill_1.dat};
          \node[ulgRed, anchor=west, hlLabel] at (axis cs: 0.25, 0.1) {R1};
      \end{scope}
      \addplot[rheoStyle] coordinates {(\fiRheoals,\fiYminals) (\fiRheoals,\fiYmaxals)};
      \draw[thick, ulgBlueGreen, -Stealth, dashed] (axis cs: \fiRheoals, 0.1) -- node[midway, hlLabel, ulgBlueGreen, anchor=south] {R2} (axis cs:0.6, 0.1);
      
    \nextgroupplot[
        xmode=log, xmin=0.4, xmax=1.0,
        ymode=log, ymin=0.01, ymax=300, 
        xtick={0.2,0.4,0.8}, xticklabels={0.2,0.4,0.8},
        yticklabels={},
        ytick={},
      ]
      \addplot[scatter, only marks, mark size=0pt, draw=none, point meta=explicit]
        coordinates {(\fiXminals,\fiYminals) [0]};
      \fiPlotFills{HPM-satk}{\fiLevelsals}
      \fiPlotLines{HPM-satk}{\fiLevelsals}
      \addplot[rheoStyle] coordinates {(0.5,\fiYminals) (0.5,\fiYmaxals)};
      \draw[thick, ulgBlueGreen, -Stealth, dashed] (axis cs: 0.5, 0.1) -- node[midway, hlLabel, ulgBlueGreen, anchor=south] {R2} (axis cs:0.6, 0.1);

    \nextgroupplot[
        xmode=log, xmin=1.0, xmax=5.0,
        ymode=log, ymin=0.9, ymax=300, 
        xtick={0.2,0.4,0.8}, xticklabels={0.2,0.4,0.8},
        yticklabels={},
        ytick={},
      ]
      \addplot[scatter, only marks, mark size=0pt, draw=none, point meta=explicit]
        coordinates {(\fiXminals,\fiYminals) [0]};
      \fiPlotFills{LIF}{\fiLevelsals}
      \fiPlotLines{LIF}{\fiLevelsals}
      \addplot[rheoStyle] coordinates {(1.5,\fiYminals) (1.5,\fiYmaxals)};
      \draw[thick, ulgBlueGreen, -Stealth, dashed] (axis cs: 1.5, 3.) -- node[midway, hlLabel, ulgBlueGreen, anchor=south] {R2} (axis cs:2.2, 3.);
  \end{groupplot}
  \node[anchor=north, align=center, fill=none]
    at ($(group c2r1.south)!0.5!(group c3r1.south)$) {Peak amplitude $w_{\mathrm{pk}}$};
    
  \node[anchor=north, text width=0.23\linewidth, align=center, fill=none]
    at ([yshift=-0.2cm]group c1r1.south) {\subcaption{\gls{cbm}\label{fig:wave-gneu}}};
  \node[anchor=north, text width=0.23\linewidth, align=center, fill=none]
    at ([yshift=-0.2cm]group c2r1.south) {\subcaption{\gls{hpm}\label{fig:wave-als}}};
    \node[anchor=north, text width=0.23\linewidth, align=center, fill=none]
    at ([yshift=-0.2cm]group c3r1.south) {\subcaption{non--excitable \gls{hpm}\label{fig:wave-hpm-satk}}}; 
    \node[anchor=north, text width=0.23\linewidth, align=center, fill=none]
    at ([yshift=-0.2cm]group c4r1.south) {\subcaption{LIF\label{fig:wave-lif}}};
\end{tikzpicture}
    \caption{\small \textbf{Transient sensitivity of the models} to excitatory waves~\eqref{eq:alpha_wave}, log-log scale. Grey levels indicate the number of spikes each model produces in response.} 
    \label{fig:alphawave_sweep_theory}
\end{figure*}

\section{Responses to transient inputs}\label{sec:excitability}

Excitability is best defined as an all-or-none sensitivity to transient inputs. This transient sensitivity underlies spikes triggered below the rheobase, including rebound spikes at the release of an inhibitory input, as well as the reliability of spike timing in the presence of parametric mismatch (\cref{fig:reliability-exp}). In this section, we show that the \gls{cbm} and the \gls{hpm} respond similarly to transient inputs, despite showing some differences near the rheobase. We then investigate how the \gls{hpm} parameters can tune its sub-rheobase sensitivity.

\subsection{Sensitivity of neuron models to transient inputs}

Inspired by~\cite{sepulchreExcitableBehaviors2017}, we apply a transient wave input to four neuron models: the \gls{cbm}~\eqref{eq:hh_model}, an excitable \gls{hpm}~\eqref{eq:hpm}, a non-excitable \gls{hpm}, and a regular \gls{lif}. The transient wave input is defined for $t\geq 0$ as
\begin{equation}
    w(t) = w_0 + w_{\mathrm{pk}}\,\frac{t}{\tau_w}\exp\!\Big(1-\frac{t}{\tau_w}\Big), \label{eq:alpha_wave}
\end{equation}
where $\tau_w$ is the timescale of the wave and $w_{\mathrm{pk}}$ is its peak amplitude (reached at $t=\tau_w$) with respect to the baseline $w_0$. The baselines are set individually for each model below its rheobase so that the neurons are initially at rest. Two protocols are performed, one with excitatory waves ($w_{\mathrm{pk}}>0$) and one with inhibitory (hyperpolarizing) waves ($w_{\mathrm{pk}}<0$). For each protocol, the number of spikes produced by each model in response to a single wave is reported in~\cref{fig:alphawave_sweep_theory_rebound,fig:alphawave_sweep_theory} as a function of the amplitude $|w_{\mathrm{pk}}|$ and the timescale $\tau_w$, both on logarithmic axes. Transposing the $x$ and $y$ axes, the figures give sensitivity curves as a function of the input's timescale range (the inverse of frequency), reminiscent of amplitude Bode plots.

\subsubsection{Response to inhibitory waves}

Let us first examine the results of the inhibitory protocol in~\cref{fig:alphawave_sweep_theory_rebound}, as they are the simplest to interpret. The non-excitable models (\gls{lif} and non-excitable \gls{hpm}) never produce rebound spikes, whereas the excitable ones (\gls{hpm} and \gls{cbm}) do, so only the latter are shown. The regions of the $(w_{\mathrm{pk}}, \tau_w)$-plane in which rebound spikes occur are qualitatively similar for the \gls{hpm} and the \gls{cbm}: rebounds are triggered after an inhibition of sufficient amplitude, and within a restricted range of timescales. This timescale range reflects the frequency selectivity to transient inputs of both models. On one hand, the mixed feedback of the \gls{cbm} behaves like a band-pass filter with a large bandwidth, since its positive feedback (the sodium activation $m_{\mathrm{Na}}$) possesses a fast, but not negligible, timescale. On the other hand, the mixed feedback of the \gls{hpm} is a high-pass filter, so its sensitivity to small timescales is larger.

\subsubsection{Response to excitatory waves} \label{sssec:resp_exec}

Let us now examine the results of the excitatory protocol in~\cref{fig:alphawave_sweep_theory}. A first striking observation is that the figures become progressively simpler, going from the \gls{cbm} to the \gls{hpm}, the non-excitable \gls{hpm}, and finally the \gls{lif}. To structure the discussion, we divide the $(w_{\mathrm{pk}}, \tau_w)$-plane of the \gls{cbm} into several qualitatively distinct regions: R1 to R4, interpret each of them, and examine whether they appear in the other models.

\begin{enumerate}[label=R\arabic*:, ref=R\arabic*, wide=\parindent]
    \item\label{item:R1} \textbf{Sub-rheobase excitability:} This region corresponds to excitatory waves that trigger a spike, despite their peak amplitude remaining below the rheobase (vertical lines in the figure). Only the excitable models (the \gls{cbm} and the \gls{hpm}) exhibit this behaviour. Similarly to rebound spikes, the region appears on a restricted range of timescales, reflecting the frequency selectivity of both models to transient inputs. In models where this region exists, spike timing is reliable (\cref{fig:reliability-exp}). This region is probably well-suited for robust spike-based coding.

    \item\label{item:R2} \textbf{Tonic spiking:} This region corresponds to inputs for which the models enter an oscillatory regime. Its interpretation is straightforward: the longer the input remains above the rheobase, the more spikes are produced. All models exhibit a version of this region. Interestingly, the purely low-pass filtering mechanism of the \gls{lif} delays significantly the first spike trigger, which explains the absence of spiking over a large region of R2 for small timescales. A similar effect can be found in the other models, but it is confined to much smaller timescales. This region is well-suited for rate-based coding.

    \item\label{item:R3} \textbf{Rest-spike bistability:} Because of accommodation, depending on the protocol, two rheobases can be found for the \gls{cbm}: the lower one is identified using step inputs, and the upper one using a very slowly rising ramp~\cite{Hill1936,AzouzGray2000,PlatkiewiczBrette2010}. We bound R1 by the lower rheobase, since spikes beyond it can lead to tonic spiking, and R2 by the upper rheobase, since this is where the resting state truly loses stability. R3 corresponds to a hysteresis effect which is absent from the other models~\cite{Izhikevich2000,LuXinRinzel2023}. Neurons exhibiting this region can better retain memory of past inputs. We conjecture that a \gls{hpm} with a second-order filter would recover this region.

    \item\label{item:R4} \textbf{Spike latency:} For slowly varying inputs, the \gls{cbm} driven slightly above its upper rheobase may still take a substantial time before producing its first spike. This explains a gap in R2, which we call R4, where the neuron is on its way to tonic spiking, but the input wave returns it to rest before the first spike is produced. We attribute this delay to the timescale of the positive feedback (sodium activation $m_{\mathrm{Na}}$). Neglecting it, as in the \gls{hpm}, removes the delay between excitation and spike production. Neurons exhibiting this region display a dynamic phase response to transient inputs. We conjecture that a \gls{hpm} with a band-pass filter of wide bandwidth would recover this region.
\end{enumerate}

\subsection{Tuning the sub-rheobase sensitivity of a \gls{hpm}}

\begin{figure}[t]
    \vspace{2pt}
%
%

\providecommand{\fiDir}{./_datas/parameters_influence}  
\providecommand{\fiIdxStart}{1}                  
\providecommand{\fiNk}{5}                        
\providecommand{\fiNp}{5}                        
\providecommand{\fiNtau}{5}                      
\providecommand{\fiNalpha}{5}                    
\providecommand{\fiMinOpacity}{0.15}             
\providecommand{\fiXmin}{0.0} \providecommand{\fiXmax}{0.9}
\providecommand{\fiYmin}{0.01} \providecommand{\fiYmax}{300}
\providecommand{\fiArrowFromK}{0.46,100}      \providecommand{\fiArrowToK}{0.27,100}
\providecommand{\fiArrowFromP}{0.10,10}       \providecommand{\fiArrowToP}{0.55,10}
\providecommand{\fiArrowFromTau}{0.2,0.05}   \providecommand{\fiArrowToTau}{0.2,10.0}
\providecommand{\fiArrowFromAlpha}{0.2,1}  \providecommand{\fiArrowToAlpha}{0.5,1}

\colorlet{parK}{ulgRed}
\colorlet{parP}{ulgBlue}
\colorlet{parTau}{ulgViolet}
\colorlet{parAlpha}{ulgGreen}
\pgfplotsset{
  paramLine/.style={thick, forget plot},
}

\newcommand{\fiParamLines}[3]{%
  \pgfmathtruncatemacro{\fiLastIdx}{\fiIdxStart+#2-1}%
  \foreach \fiI in {\fiIdxStart,...,\fiLastIdx}{%
    \pgfmathsetmacro{\fiOp}{\fiMinOpacity+(1-\fiMinOpacity)*(\fiI-\fiIdxStart)/max(#2-1,1)}%
    \edef\fiTmp{\noexpand\addplot[paramLine, #3, opacity=\fiOp]
      table[x=x, y=y] {\fiDir/#1_line_\fiI.dat};}\fiTmp
  }%
}

\newcommand{\fiPanel}[3]{%
  \fiParamLines{#1}{#2}{#3}%
}

\begin{tikzpicture}
  \begin{groupplot}[
      group style={
        group size=2 by 2,
        horizontal sep=0.3cm, vertical sep=0.7cm,
        xticklabels at=edge bottom, yticklabels at=edge left,
      },
      scale only axis,
      width=0.4\linewidth, height=3.0cm,
      xmode=normal, ymode=log,
      xmin=\fiXmin, xmax=\fiXmax, ymin=\fiYmin, ymax=\fiYmax,
      xtick={0.1,0.45,0.8}, xticklabels={0.1,0.45,0.8},
      ytick={0.01,1,100},
      title style={font=\normalsize, yshift=-0.15cm},
      tick label style={font=\scriptsize},
    ]
    \nextgroupplot[title={(a) $\tau$: $1\!\to\!100$}]
      \fiPanel{taus}{\fiNtau}{parTau}
      \draw[black!50, -Latex, thick] (axis cs:\fiArrowFromTau) to[bend left=30] (axis cs:\fiArrowToTau);
    \node[draw, anchor=north west, font=\tiny, inner sep=0.1cm, fill=gray!20, align=left] at (axis cs:0.5,5) {$
         k\!=\!2$\\
         $p\!=\!2.7$\\
         $\alpha\!=\!0.01$};
    \nextgroupplot[title={(b) $\alpha$: $0.01\!\to\!0.5$}]
      \fiPanel{alpha}{\fiNalpha}{parAlpha}
      \draw[black!50, -Latex, thick] (axis cs:\fiArrowToAlpha) to[bend left=20] (axis cs:\fiArrowFromAlpha);
    \node[draw, anchor=north west, font=\tiny, inner sep=0.1cm, fill=gray!20, align=left] at (axis cs:0.55,5) {$
         k\!=\!2$\\
         $p\!=\!25$\\
         $\tau\!=\!10\si{\milli\second}$};
    \nextgroupplot[title={(c) $p$: $1\!\to\!50$}]
      \fiPanel{p}{\fiNp}{parP}
      \draw[black!50, -Latex, thick] (axis cs:\fiArrowFromP) to[bend left=30] (axis cs:\fiArrowToP);
    \node[draw, anchor=north west, font=\tiny, inner sep=0.1cm, fill=gray!20, align=left] at (axis cs:0.55,5) {$
         k\!=\!2$\\
         $\tau\!=\!10\si{\milli\second}$\\
         $\alpha\!=\!0.01$};
    \nextgroupplot[title={(d) $k$: $1.2\!\to\!4$}]
      \fiPanel{k}{\fiNk}{parK}
    \draw[black!50, -Latex, thick] (axis cs:0.4, 150.0) ..controls (axis cs:0.5, 120.0) and (axis cs:0.6, 100.0).. (axis cs:0.22, 90.0);
    \node[draw, anchor=north west, font=\tiny, inner sep=0.1cm, fill=gray!20, align=left] at (axis cs:0.5,10) {$
         p\!=\!2.7$\\
         $\tau\!=\!10\si{\milli\second}$\\
         $\alpha\!=\!0.01$};
  \end{groupplot}
  \node[anchor=north, inner sep=0.0] at ($(group c1r2.south)!0.5!(group c2r2.south)+(0,-0.5cm)$)
      {Peak amplitude $w_{\mathrm{pk}}$};
  \node[anchor=south, rotate=90] at ($(group c1r1.west)!0.5!(group c1r2.west)+(-0.8cm,0)$)
      {Timescale $\tau_w$};
  \coordinate (bbsw) at (current bounding box.south west);
  \coordinate (bbne) at (current bounding box.north east);
  \pgfresetboundingbox
  \path[use as bounding box] ($(bbsw)+(0,0.4cm)$) rectangle (bbne);
\end{tikzpicture}
    \vspace{0.5mm}
    \caption{\small \textbf{Parameter influence on the sub-rheobase response of the \gls{hpm}.} Boundary of the spiking region in the $(w_{\mathrm{pk}},\tau_w)$-plane. Spikes occur to the right of each curve. Each panel increases one parameter; the others are fixed.}
    
    \label{fig:parameters_on_alpha-input}
\end{figure}
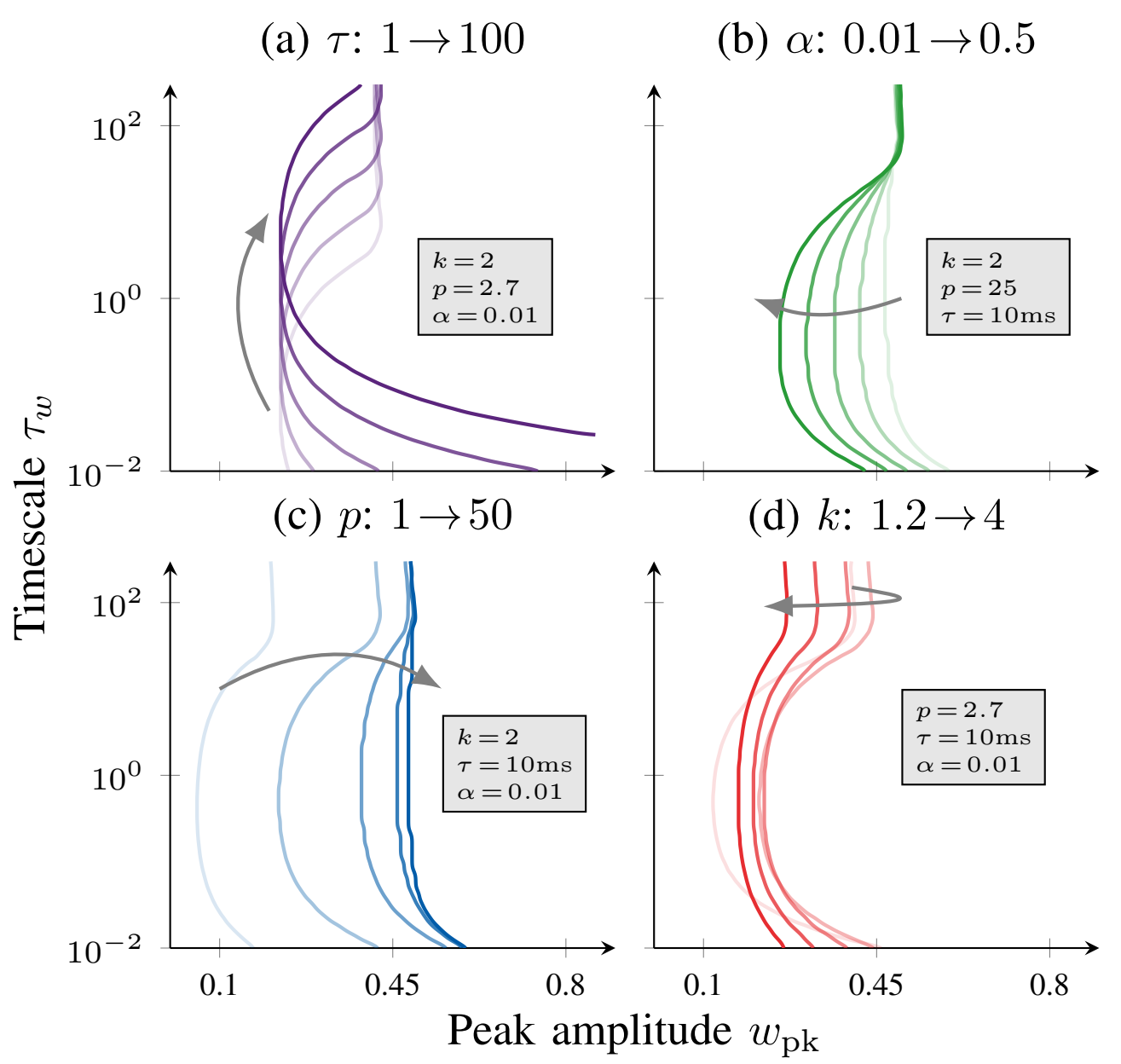

\Cref{fig:parameters_on_alpha-input} shows how the boundary of the R1 spiking region in the $(w_{\mathrm{pk}},\tau_w)$-plane changes when each parameter of the \gls{hpm} is varied while the others are held fixed. The four parameters play distinct roles:
\begin{itemize}
    \item $\tau$ shifts the spiking boundary along the timescale axis, leaving both the amplitude range of R1 and the rheobase unchanged. This is expected, since $\tau$ sets the frequency selectivity of the \gls{hpm}.
    \item $\alpha$ changes the amplitude range of R1, leaving the timescale sensitivity unchanged. The rheobase stays nearly constant at large $p$, but can shift at smaller $p$. This is consistent with $\alpha$ tuning the loop gain of the \gls{hpm} in its excitable sub-rheobase regime.
    \item $p$ changes both the rheobase and the amplitude range of R1, since the rheobase lies at the corner of the sigmoid while the sub-rheobase sensitivity is set by the gain below it, thus only leaving the timescale sensitivity unchanged. For $\alpha= 0$, the excitability of the \gls{hpm} relies entirely on the corner of the sigmoid not being too sharp, i.e., on $p$ remaining small.
    \item $k$, unlike the other parameters, has a non-monotonic effect on the spiking boundary: as $k$ increases, the rheobase first increases and then decreases, while R1 flattens. This effect is harder to interpret, as this gain influences the sharpness of the sigmoid corner in a way not captured by $p$. Since $k$ indirectly affects the entire spiking boundary, we recommend keeping it fixed.
\end{itemize}
In short, $\tau$ sets the timescale at which the \gls{hpm} responds, $\alpha$ and $p$ mainly set how far below the rheobase it responds, and $p$ and $k$ set the rheobase itself. More generally, we expect the R1 spiking boundary of \glspl{hpm} to be shifted along the amplitude axis by the sigmoid parameters, which set the amplitude sensitivity, and along the timescale axis by the frequency-selector parameters, which set the timescale sensitivity. This is an important advantage, as it allows \glspl{hpm} to be tuned intuitively by acting on each sensitivity separately.

\section{Responses to constant inputs}\label{sec:tonic_regime}

\begin{figure}[t]
    \centering
    \vspace{2pt}
\colorlet{fiData}{ulgBlueGreen}
\colorlet{leftbranch}{ulgRed}
\colorlet{rightbranch}{ulgBlueGreen}
\pgfplotsset{
  fiPoints/.style={only marks, mark=*, mark size=1.2pt, fiData},
  fiRheoLine/.style={dashed, thick, fiData},
}

\begin{tikzpicture}
  \begin{axis}[
      width=8cm, height=5.5cm,
      xtick={-50,0,50,150,250,...,450},
      xlabel={Current $I_{\mathrm{ext}}$ ($\mu$A/cm$^2$)},
      ylabel={Frequency (Hz)},
      xmin=-10, xmax=500,
      ymin=-0.5,
    ]
    \addplot[fiData, thick, restrict x to domain=0.0:11.2, leftbranch] table[x=I, y=f] {_datas/rheo_data.dat};
    \addplot[fiData, thick, restrict x to domain=11.2:500, leftbranch] table[x=I, y=f] {_datas/rheo_data.dat};
    \addplot[fiRheoLine, leftbranch] coordinates {(11.207,0) (11.207,46.94)};

    \draw[black!70, thick] (axis cs:7.0, 0.0) rectangle (axis cs:18,60);
    \coordinate (zoomBR) at (axis cs:18.0,0.0);
    \coordinate (zoomTR) at (axis cs:18.0,60);
    \coordinate (insetpos) at (axis cs:350,20.);   
    
  \end{axis}
\begin{axis}[
    name=zoom,
    at={(insetpos)}, anchor=south,
    width=4cm, height=3.5cm,
    xmin=7.0, xmax=18.0, ymin=0.0, ymax=60.0,
    no markers,
    axis lines=box,
    xtick={10.0, 13.0, 15.0}, ytick={0.0, 30.0, 60.0},
    tick label style={font=\tiny},
    ticklabel shift=-.1cm,
    every axis plot/.append style={thick},
    axis background/.style={fill=white},
    axis line style={black!50},
]

    \addplot[fiData, thick, restrict x to domain=11.2:90, leftbranch] table[x=I, y=f] {_datas/rheo_data.dat};
    \addplot[fiRheoLine, leftbranch] coordinates {(11.207,0) (11.207,46.94)};
    \addplot[fiData, very thick, leftbranch] coordinates {(0,0) (11.207,0)};
    \addplot[fiRheoLine, rightbranch] coordinates {(13.01,0) (13.01,52.06)};
    \addplot[fiData, very thick, rightbranch] coordinates {(11.207,0) (13.01,0)};
    \node[anchor=east, fiData, font=\tiny, leftbranch] at (axis cs:11.,10) {\rheoLow};
    \node[anchor=west, rightbranch, font=\tiny] at (axis cs:13,10) {\rheoUp};
\end{axis}
\draw[black!70, dashed] (zoomBR) -- (zoom.south west);
\draw[black!70, dashed] (zoomTR) -- (zoom.north west);
\end{tikzpicture}
    \caption{\small \textbf{$f$-$I$ curve for the \gls{cbm}.} The inset zooms on the neighbourhood of the rheobases, indicated by dashed vertical lines: the lower rheobase \rheoLow{} (red) and the upper rheobase \rheoUp{} (teal).}
    \label{fig:cbm_theo_fi}
\end{figure}
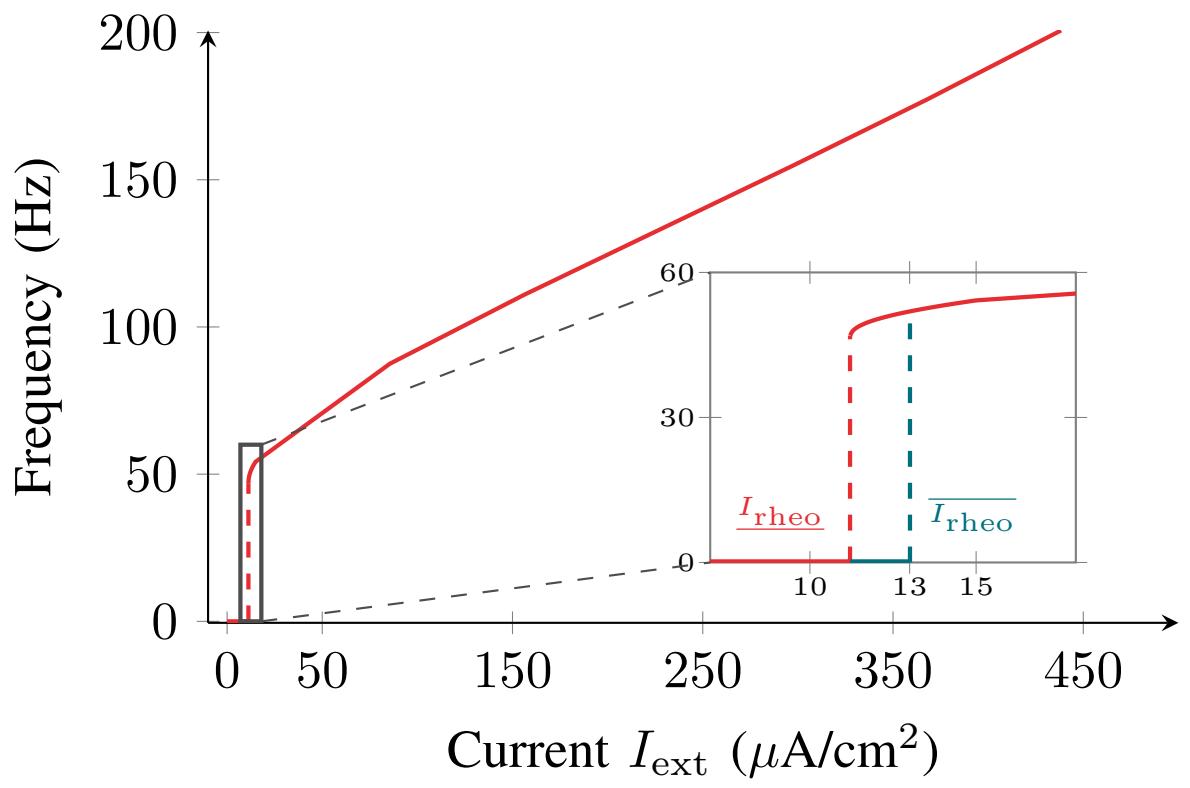

As discussed previously, for constant inputs, both the \gls{cbm} and the \gls{hpm} fire tonically above their rheobase. In this section, we aim to fit the $f$-$I$ curve of the \gls{hpm} to that of a given  \gls{cbm}, such as the one illustrated in~\cref{fig:cbm_theo_fi}. Fitting reduced neuron models to detailed ones has a long history~\cite{Jolivet2004,BretteGerstner2005,DrionSepulchre2021}, but our goal here is narrower: we only aim to evaluate how easily the tonic spiking regime of a \gls{hpm} can be made to coincide with that of a \gls{cbm}.
A first difficulty is that the $f$-$I$ curves of the two models differ qualitatively in shape. First, as mentioned in~\cref{sssec:resp_exec}, the \gls{cbm} has two rheobases, which we now denote \rheoLow{} and \rheoUp{}, between which it is bistable. However, the first-order \gls{hpm} is monostable and has a single rheobase. Moreover, the $f$-$I$ curve of the \gls{cbm} increases first concavely and then almost linearly, whereas that of the \gls{hpm} is typically more bell-shaped~\cite{CHRI-JE-ECC}.

To overcome this difficulty, we assume that the bistable window of the \gls{cbm} is sufficiently narrow for a single rheobase in the \gls{hpm} to be a reasonable approximation, and we only use the first half of the \gls{hpm} $f$-$I$ curve, where the frequency $f$ increases with $u$. This latter point is achieved through an affine parametrization of the input, such that
\begin{equation}
    u=\varphi(I_\mathrm{ext}):=\frac{1}{\rho}(I_\text{ext}-I_0),\qquad \rho>0,
    \label{eq:affine_map}
\end{equation}
where $I_0$ is the current mapped to $u=0$, the maximum spiking frequency of the \gls{hpm}. The fit is performed on $[I_r, I_0]$, where $I_r$ is the rheobase to match. The rheobase $u_{\mathrm{rheo}}(k,p,\alpha)$ of the \gls{hpm}~\cite{CHRI-JE-ECC} fixes the scaling factor $\rho$: requiring $\varphi(I_r)=-u_{\mathrm{rheo}}$ imposes
\begin{equation}
    \rho=(I_0-I_r)/u_{\mathrm{rheo}}(k,p,\alpha),
    \label{eq:rho}
\end{equation}
so $\rho$ is not a free parameter. Given that the \gls{cbm} has two rheobases and the \gls{hpm} one, we take the midpoint $I_r=\tfrac12\big(\underline{I_{rheo}}+\overline{I_{rheo}}\big)$. The remaining parameters $\theta=(k,p,\tau,\alpha,I_0)$ are obtained by solving the nonlinear least-squares problem
\begin{equation}
    \min_{\theta} \sum_{i=1}^{N}\Big[f_{\mathrm{CBM}}(I_i)-f_{\mathrm{HPM}}\big(\varphi(I_i);\theta\big)\Big]^2,\!\!\quad I_i\in[I_r,I_0],
    \label{eq:lsq}
\end{equation}
where $f_{\mathrm{CBM}}$ is the frequency measured on the tonic-spiking branch and $f_{\mathrm{HPM}}$ the analytical frequency of the \gls{hpm}~\cite{CHRI-JE-ECC}. We enforce $k>1$ and $p,\alpha>0$ so the \gls{hpm} remains excitable. The expression of $f_{\mathrm{HPM}}$ contains a root-finding step, which is differentiable through the implicit function theorem~\cite{Blondel2022}. This gives the Jacobian of the residuals, and we solve \eqref{eq:lsq} with the Levenberg-Marquardt algorithm~\cite{Levenberg1944,Marquardt1963}. We recall that the implementation details are available in the GitHub repository mentioned earlier.

\begin{figure}[t]
    \centering
    \vspace{2pt}

\colorlet{fitData}{black!70}
\colorlet{fitHPM}{ulgOrange}
\colorlet{rheoLow}{ulgRed}
\colorlet{rheoUp}{ulgBlueGreen}
\colorlet{rheoMean}{ulgOrange}
\pgfplotsset{
  fitPoints/.style={only marks, mark=*, mark size=1.6pt, fitData},
  fitCurve/.style={very thick, fitHPM},
  rheoVline/.style={thick, opacity=0.45},
}

\begin{tikzpicture}
  \begin{axis}[
      width=\linewidth, height=3.5cm,
      xlabel={$I_{\mathrm{ext}}$ ($\mu$A/cm$^2$)},
      ylabel={Frequency (Hz)},
      xmin=8, xmax=99,
      ymin=0, ymax=99,
      legend pos=south east,
      legend cell align=left,
      legend style={draw=none, fill=none},
      legend columns=2
    ]

    \addplot[rheoVline, rheoLow]  coordinates {(11.207,0)  (11.207,100)};
    \addlegendentry{\rheoLow}
    \addplot[rheoVline, rheoUp]   coordinates {(13.01,0)   (13.01,100)};
    \addlegendentry{\rheoUp}
    \addplot[rheoVline, rheoMean, dashed] coordinates {(12.1085,0) (12.1085,100)};
    \addlegendentry{Avg $I_{\mathrm{rheo}}$}

    \addplot[fitCurve, restrict x to domain=0:12.13, forget plot]
      table[x=I, y=f_fit] {_datas/Fitting_HPM_dat/if_curve.dat};
    \addplot[fitCurve, restrict x to domain=12.13:100]
      table[x=I, y=f_fit] {_datas/Fitting_HPM_dat/if_curve.dat};
    \addlegendentry{\gls{hpm} fit}

    \addplot[fitPoints] table[x=I, y=f_data] {_datas/Fitting_HPM_dat/fit_points.dat};
    \addlegendentry{data points}

    \node[anchor=north west, draw=black!50, fill=white, inner sep=2pt, font=\footnotesize]
      at (axis cs:20,96) {RMSE $=0.29$\,Hz};
  \end{axis}
\end{tikzpicture}
    \caption{\small \textbf{Fitted $f$-$I$ curve.} \gls{hpm} (\textcolor{ulgOrange}{orange}) fitted to the frequencies of the \gls{cbm} (dots) on $[I_r,I_0]$ with an affine input map. Vertical lines: \rheoLow, \rheoUp{} and their mean $I_r$, imposed on the \gls{hpm}. Fitted parameters: $k=1.13$, $p=5.61$, $\alpha=0.55$, $\tau=31.5\si{\milli\second}$, $I_0=221.4\si{\micro\ampere/\centi\meter^2}$, $\rho=303.5$.}
    \label{fig:fitted-fi-curve}
\end{figure}
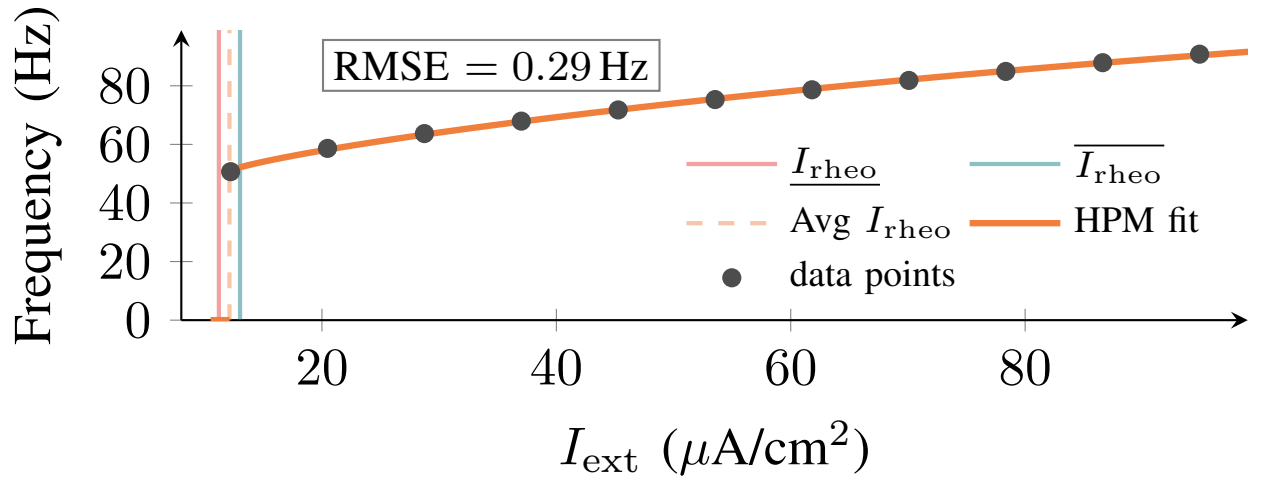

\Cref{fig:fitted-fi-curve} shows the result for $N=11$ equally spaced currents. The \gls{hpm} reproduces the onset frequency and the concave rise with a root-mean-square error of \SI{0.29}{\hertz}, and below \SI{0.5}{\hertz} at every point. This suggests that a simple affine remapping of the \gls{hpm} input is sufficient to match its tonic spiking regime to that of a \gls{cbm}.



\section{Conclusions and perspectives}\label{sec:conclusion}

In this work, we investigated how expressive and easily tunable the first-order \gls{hpm} is, using a simple Hodgkin-Huxley-type \gls{cbm} as a reference. Below the rheobase, we showed that the \gls{hpm} reproduces the main excitable behaviours of the \gls{cbm}. Subjected to transient wave inputs, both models exhibit sub-rheobase excitability within a restricted range of timescales, thus making their spike timing reliable. These behaviours are absent from non-excitable models such as \gls{lif} neurons. Moreover, the parameters of the \gls{hpm} have a relatively clear and decoupled influence on its input sensitivity, which makes the model particularly intuitive to tune. Above the rheobase, we showed that, despite some qualitative differences, the $f$-$I$ curve of the \gls{cbm} can be matched by that of an \gls{hpm} using a simple affine parametrization of the \gls{hpm} input. The remaining differences between the two models concern mechanisms of the \gls{cbm} located near its rheobase. The \gls{cbm} exhibits a bistability region, and it can delay its first spike after a slow excitation. In contrast, the first-order \gls{hpm} is monostable and generally fires immediately after excitation. We conjecture that these behaviours can be recovered within the \gls{hpm} framework, with a second-order high-pass filter for bistability and a band-pass filter of wide bandwidth for spike latency. 

Overall, these results indicate that even in its simplest form, the \gls{hpm} captures the essential excitable and tonic spiking behaviours of a conductance-based neuron, while remaining analytically tractable and intuitive to tune, making it a promising candidate for neuromorphic applications.

Significant work remains to facilitate the theoretical design, numerical simulation, and hardware implementation of the \gls{hpm}, particularly for frequency selectors more complex than the first-order high-pass filter investigated here.
\bibliographystyle{ieeetr}
\bibliography{references}
\end{document}